\documentstyle[12pt,a4,epsf]{article}
\begin{document} 
\bibliographystyle{unsrt} 
\baselineskip= 18pt 
\pagenumbering{arabic} 
\pagestyle{plain} 
\def \bit{\begin{itemize}}
\def \eit{\end{itemize}}
\def \half{{1\over 2}}
\def \OO{\Omega}
\def \aa{\alpha}
\def \bk{{\bf k}}
\def \bkp{{\bf k'}}
\def \bqp{{\bf q'}}
\def \bq {{\bf q}}
\def \bpp{{\bf p'}}
\def \bp {{\bf p}}
\def \EE{\Bbb E}
\def \EEx{\Bbb E^x}
\def \EEo{\Bbb E^0}
\def \LL{\Lambda}
\def \PP{\Bbb P^o}
\def \rr{\rho}
\def \SS{\Sigma}
\def \ss{\sigma}
\def \ll{\lambda}
\def \dd{\delta} \def \ee{\epsilon} \def \ww{\omega}
\def \bww{\mbox{\boldmath $\omega$}}
\def \DD{\Delta}
\def \DDt{\tilde {\Delta}}
\def \kr{\kappa\lb \LL\rb}
\def \PPx{\Bbb P^{x}}
\def \gg{\gamma}
\def \GG{\Gamma}
\def \kk{\kappa}
\def \tt{\theta}
\def \lb{\left(}
\def \la{\langle}
\def \ra{\rangle}
\def \rb{\right)}
\def \prt{\tilde p}
\def \pt{\tilde {\phi}}
\def \bb{\beta}
\def \hal{{1\over 2}\nabla ^2}
\def \bg{{\bf g}}
\def \bx{{\bf x}}
\def \bu{{\bf u}}
\def \bv{{\bf v}}
\def \by{{\bf y}}
\def \bw{{\bf w}}
\def \bU{{\bf U}}
\def \hag{{1\over 2}\nabla}
\def \beq{\begin{equation}}
\def \eeq{\end{equation}}
\def \bea{\begin{eqnarray}}
\def \eea{\end{eqnarray}}
\def \cosech{\hbox{cosech}}
\def \fr{\frac}
\def \d{\partial}
\def \Gc{{\cal G}}
\def \dmq{\fr{d^3\bq}{(2\pi)^3}}
\def \dmqp{\fr{d^3\bq'}{(2\pi)^3}}
\def \dmk{\fr{d^3\bk}{(2\pi)^3}}
\def \dmkp{\fr{d^3\bk'}{(2\pi)^3}}
\def \dmp{\fr{d^3\bp}{(2\pi)^3}}
\def \dmpp{\fr{d^3\bp'}{(2\pi)^3}}
\def \Tr{{\mbox{Tr}}}

\title{Renormalization of Drift and Diffusivity in Random Gradient Flows}
\author{D S Dean$^{\dag}$, I T Drummond and R R Horgan  \\
        Department of Applied Mathematics and Theoretical Physics \\
        University of Cambridge \\
        Silver St \\
        Cambridge, England CB3 9EW\\~~\\~~\\
        ${~^{\dag}}$Dipartimento di Fisica and INFN\\
        Univesit\'a di Roma La Sapienza \\
        P. A. Moro 2, 00185 Roma, Italy}
\maketitle

\begin{abstract}

We investigate the relationship between the effective diffusivity
and effective drift of a particle moving in a random medium. The velocity 
of the particle combines a white noise
diffusion process with a local drift term that depends linearly on 
the gradient of a gaussian random field with homogeneous statistics.
The theoretical analysis is confirmed by numerical simulation. 

For the purely isotropic case
the simulation, which measures the effective drift directly in a constant gradient
background field, confirms the result previously obtained theoretically, 
that the effective diffusivity and effective drift are renormalized by the same
factor from their local values. For this isotropic case we provide an intuitive
explanation, based on a {\it spatial} average of local drift, for the 
renormalization of the effective drift parameter relative to its local value.

We also investigate situations in which the isotropy is broken by the tensorial
relationship of the local drift to the gradient of the random field. 
We find that the numerical simulation
confirms a relatively simple renormalization group calculation for the effective
diffusivity and drift tensors.

\end{abstract}
\vfill
DAMTP-96-34
\pagebreak
 
\section{Introduction}

A problem of great interest concerns the 
computation of effective parameters for a diffusion process that combines molecular
diffusion with a drift term linearly dependent on the gradient of
a random scalar field with spatially homogeneous statistics. In the case where the 
system, including the statistical properties of the random 
field, is isotropic it has been shown, on the basis of plausible
but not completely established assumptions, that the effective long range diffusivity
is accurately predicted by a renormalization
group calculation (RGC) \cite{mythesis,dc,USDIF1}. The result, which is 
embodied in a very simple formula,
is confirmed by numerical simulation to high accuracy. The result is also consistent 
with the straightforward perturbation expansion to two-loop order but remains accurate beyond
the applicability of the expansion to this order \cite{dc,USDIF2}. 

An important parameter relevant to the long time 
behaviour of the system, that has not generally been
studied, is the value of the effective drift parameter, or tensor, that determines
the average drift of particles in an imposed large-scale field gradient. A feature of the
isotropic case, respected by the RGC, is that this effective drift coefficient is
renormalized relative to its local value by the same
factor as the effective diffusivity is renormalized relative to the molecular 
diffusivity \cite{mythesis,dc,USDIF1,USDIF2}. By including an appropriate constant 
drift term in the original molecular diffusion process we confirm, in this paper, 
directly by numerical simulation,
the equality of the two renormalization factors. We also give an intuitive
explanation in terms of spatial averages, of why one should expect that
the drift coefficient be renormalized.

Isotropy can break down for three reasons. First, the statistics of the random field
may not be isotropic. We have investigated this in a previous paper \cite{USDIF3}
and shown that, although the
predictions of the RGC are still reasonably accurate 
there are emerging disparities with the results of the numerical simulation. 
Such an outcome is to be expected since the results of the RGC 
were, in this case, shown to differ from
straightforward perturbation theory at two-loop order. 
Second, the molecular diffusion
tensor may be non-isotropic and third, the same may be true of the local drift tensor.
It is this third possibility that we investigate in this paper by numerical simulation 
and compare with the predictions of the appropriate RGC. The results actually agree
rather well suggesting that the isotropy of the random field statistics is
an important condition for the success of the RGC.

\section{Diffusion with Background Drift}

It is convenient for the purpose of discussing anisotropy to consider the
most general diffusion equation of the class
in which we are interested. It has the form
\beq
\fr{\d~P}{\d~t}=\d_i(\kk^0_{ij}\d_j-\ll^0_{ij}\d_j\phi(\bx)-u_i)P~~.
\eeq
Here $P$ is the probability density of a particle moving according to the equation
\beq
\dot{x}_i=\ll^0_{ij}\d_j\phi(\bx)+u_i+w_i(t)~~,
\eeq
where the white noise term has zero mean and the correlation function
\beq
\la w_i(t)w_j(t')\ra=2\kk^0_{ij}\delta(t-t')~~,
\eeq
and $u_i$ is a constant drift term. In appropriate units we would expect
the drift to be given by
\beq
u_i=\ll^0_{ij}g_j~~,
\eeq
where $g_i$ is a uniform gradient field.
 
The equation for the static Green's function corresponding to a unit source at $\bx'$
is
\beq
\d_i\left(\kk^0_{ij}\d_j-\ll^0_{ij}\d_j\phi(\bx)-u_i\right)G(\bx,\bx')=-\dd(\bx-\bx')~~.
\eeq
The effective Green's function obtained after averaging over the random ensemble of 
flows we denote by $\Gc(\bx-\bx')$~. It is simpler to study this function in
terms of its Fourier transform. It has the form
\beq
\tilde{\Gc}(\bk)=[\kk^0_{mn}k_mk_n-\SS(\bk)+iu_jW_j(\bk)]^{-1}~~.
\eeq
At small $\bk$ the irreducible two-point function $\SS(\bk)$ satifies
\beq
\SS(\bk)\sim\ss_{ij}k_ik_j~~,
\eeq
with the result that the effective long range diffusivity is
\beq
\kk_{ij}=\kk^0_{ij}-\ss_{ij}~~,
\eeq
and 
\beq
W_j(\bk)\sim k_i\mu_{ij}.
\eeq
for some coefficient $\mu_{ij}$ that we now evaluate.
We show below that  
\beq
W_{i}=[(\ll^0)^{-1}]_{im}V_m(0,\bk)~~,
\eeq	
where $V_m(\bq,\bk)$ is the (Fourier transform of the) vertex function that measures the influence of a weak external 
field on $\Gc(\bx,\bx')$~. It is calculated below.
For small $\bk$ we have
\beq
V_m(0,\bk)\sim k_n\ll_{nm}~~,
\eeq
where $\ll_{mn}$ is the effective coupling referred to in the introduction.
It follows that for small $\bk$
\beq
\tilde{\Gc}(\bk)\sim [\kk_{mn}k_mk_n+ik_m\ll_{mn}g_n]^{-1}~~.
\eeq
The interpretation of this result is that the effective drift is
\beq
u^{\mbox{e}}_m=\ll_{mn}g_n~~.
\eeq
The measurement of the effective drift in a given external field allows us to extract 
the effective coupling $\ll_{ij}$ from the simulation.

For the purposes of simulation we assumed that
\beq
\DD(\bx-\bx')=\int\fr{d^3\bq}{(2\pi)^3}D(\bq)e^{i\bq.(\bx-\bx')}~~,
\eeq
with
\beq
D(\bq)=\fr{(2\pi)^{3\over 2}}{k_0^3}e^{-q^2/2k_0^2}~~.
\label{spectral1}
\eeq
With this normalization 
\beq
\la(\phi(\bx))^2\ra=\DD(0)=1~~.
\eeq

\section{Graphical Rules for Perturbation Theory}

The Feynman rules for the diagrammatic perturbation expansion are essentially
tha same as in the isotropic case. We have
\bit
\item[(i)] The sum of the inwardly flowing wave-vectors at each vertex is zero.
\item[(ii)] Each solid line carries a factor of $\displaystyle{{1\over{\kk^0_{ij}k_ik_j}}}$.
\item[(iii)] Each loop wave-vector $\bq$ is integrated with a factor $\displaystyle{{d^3\bq\over
 (2\pi)^3}}$.
\item[(iv)] Each vertex of the form of Fig. 1 carries a factor
$q_i\ll^0_{ij}\lb\bk
 +\bq\rb_j $.
\item[(v)] Each vertex of the form of Fig. 2 carries a factor
$-iu_jk_j $.
\item[(vi)] Each dashed line carries a factor $D\lb \bq \rb$.
\eit

\section{One-Loop Contributions}

The one-loop contribution to $\SS(\bk)$ is associated with Fig. 3~.
It is, according to the Feynman rules,
\beq
\SS^{(1)}(\bk)
 =-\int\dmq D(\bq)\fr{(\bk+\bq)_i\ll^0_{ij}q_j~k_m\ll^0_{mn}q_n}{(\bk+\bq)_r\kk^0_{rs}(\bk+\bq)_s}~~.
\label{eq:HH}
\eeq

The one-loop contribution to $u_iW_i(\bk)$ is associated with Fig. 4~.
The Feynman rules imply that it has the form
\beq
u_iW_i(\bk)=-\int\dmq\fr{(\bk+\bq)_i\ll^0_{ij}q_j~(-iu_j(\bk+\bq)_j)~k_m\ll^0_{mn}q_n}
                                    {((\bk+\bq)_r\kk^0_{rs}(\bk+\bq)_s)^2}~~.
\eeq
If we compare this result with that for the general vertex at one-loop
associated with Fig. 5, namely
\beq
V_i(\bq,\bk')=-\int\dmq\fr{(\bk+\bq)_j\ll^0_{jl}q_l~\ll^0_{ip}(\bk'+\bq)_p~k'_m\ll^0_{mn}q_n}
                     {(\bk+\bq)_r\kk^0_{rs}(\bk+\bq)_s(\bk'+\bq)_u\kk^0_{uv}(\bk'+\bq)_v}~~.
\eeq
we see immediately that
\beq
V_i(0,\bk)=\ll^0_{ij}W_j(\bk)~~.
\eeq
This result is easily generalized to all orders in perturbation theory.
It is equivalent to the relation used in a previous paper \cite{USDIF2}.    

\section{Spatial Averages and Mean Drift}

Some insight into the renormalization of the drift coefficient
in the isotropic case ($\ll^{(0)}_{ij}=\ll_0\delta_{ij}$ and $\kk^{(0)}_{ij}=\kk_0\delta_{ij}$)
can be obtained by considering a situation in which we allow a large cloud of particles 
to equilibriate in a large cubical sample with periodic boundary conditions. We then disturb the 
situation by imposing a uniform background drift field ${\bf g}$ causing a mean drift in the 
particle cloud. The average velocity of the particles is then renormalized relative to the
mean local drift because of the lack of uniformity in the background probability 
distribution adopted by the particles.

In the absence of a background drift field the particles adopt a density distribution $P_0(\bx)$
that yields zero particle current. That is
\beq
{\bf J}=(\kk_0\nabla-\ll_0\nabla\phi(\bx))P_0(\bx)=0~~.
\eeq
The solution of this equation is
\beq
P_0(\bx)=N\exp\{\fr{\ll_0}{\kk_0}\phi(\bx)\}~~,
\eeq
where we will choose $N$ so that $P_0(\bx)$ can be viewed as a probability distribution.
\beq
\int_{\mbox{cube}}d^3\bx P_0(\bx)=1~~.
\eeq
For a very large cube it is acceptable to enforce this equation as an ensemble average.
This allows us to evaluate $N$ as
\beq
N=\left(V\exp\left\{\fr{1}{2}\left(\fr{\ll_0}{\kk_0}\right)^2\Delta(0)\right\}\right)^{-1}~~,
\eeq
where $V$ is the volume of the cube.

When an external constant drift ${\bf g}$ is applied the system equilibriates
with a new distribution $P_0+P_1$ that satisfies
\beq
\nabla.(\kk_0\nabla-\ll_0\nabla\phi(\bx)-\ll_0{\bf g})(P_0(\bx)+P_1(\bx))=0~~.
\eeq
It turns out that no change of normalization is required.
This leads to the solution, correct to $O({\bf g})$, for $P_1$ 
\beq
P_1(\bx)=-\ll_0\int d^3\bx'G(\bx,\bx')\nabla P_0(\bx').{\bf g}~~.
\eeq
where $G(\bx,\bx')$ is the Green's function for the problem without drift.

At a point $\bx$ in a given sample of the medium the velocity of a particle
is $\ll_0(\nabla\phi(\bx)+{\bf g})$ after averaging over molecular diffusion
effects. The the drift $\bu$ of particles in the steady state situation with drift
can be obtained
as a spatial average of this velocity over the large cubical sample.
\beq
\bu=\ll_0\int d^3\bx(\nabla\phi(\bx)+{\bf g})(P_0(\bx)+P_1(\bx))
\eeq
To $O({\bf g})$ we have
\beq
u_i=\ll_0[\delta_{ij}-\ll_0\int d^3\bx d^3\bx'(\d_i\phi(\bx))G(\bx,\bx')\d_jP_0(\bx')]g_j~~.
\eeq
In fact it is reasonable now to take the ensemble average of this quantity, with the result
\beq
u_i=\ll_0[\delta_{ij}-\ll_0\int d^3\bx d^3\bx'\la(\d_i\phi(\bx))G(\bx,\bx')\d_jP_0(\bx')\ra]g_j~~.
\eeq
To lowest (non-trivial) order in $\ll_0$ we can replace $G(\bx,\bx')$ by $G_0(\bx-\bx')$
and use the identity
\beq
\la\phi(\bx)P_0(\bx')\ra=\fr{\ll_0}{\kk_0}\Delta(\bx-\bx')/V~~,
\eeq
to obtain
\beq
u_i=\ll_0[\delta_{ij}-\fr{\ll_0^2}{\kk_0}\int d^3\bx G_0(\bx)\d_i\d_j'\Delta(\bx)]g_j~~.
\eeq
Expressing this in Fourier space
we have
\beq
u_i=\ll_0[\delta_{ij}-\fr{\ll_0^2}{\kk_0^2}\int \fr{d^3\bq}{(2\pi)^3}D(q)\fr{q_iq_j}{q^2}]g_j~~.
\eeq
Taking into account the isotropy of the statistical ensemble that we assume in this case,
we see that $\bu=\ll_{\mbox{e}}{\bf g}$
where
\beq
\ll_{\mbox{e}}=\ll_0\left(1-\fr{1}{3}\fr{\ll_0^2}{\kk_0^2}\Delta(0)\right)
\eeq
This is identical with the one-loop perturbation result of previous work 
\cite{mythesis,dc,USDIF1,USDIF2}.
A careful analysis of the two-loop diagrams confirms the result to this order.
This of course, is consistent with the RGC result 
\beq
\ll_{\mbox{e}}=\ll_0\exp\left\{-\fr{1}{3}\fr{\ll_0^2}{\kk_0^2}\Delta(0)\right\}~~,
\eeq
discovered previously.
From this point of view then the origin of the renormalization of the effective drift
term comes about 
because in a steady state the particles adopt, in a given sample, 
a non-uniform distribution, appropriate to the sample, and the resulting spatial
average of the local drift is modified relative to the local ensemble average which
yields the local mean value. This should be contrasted with the situation in 
incompressible flow where the steady state distribution of particles inevitably 
remains uniform leading to a spatial average that coincides with the ensemble
average and no renormalization of the drift coefficient.

\section{Simulation of Drift in the Isotropic Case}

In Table 1 we exhibit the results of measuring both $\kk_{\mbox{e}}$ and $\ll_{\mbox{e}}$,
for an isotropic situation, over a range of values of the disorder 
parameter $\ll_0/\kk_0$~. We assume $\kk_0=1$ and $k_0=1$ throughout. The simulation,
\begin{table}
\center{
\begin{tabular}{|r|r|c|c|c|}\hline
$N$&$\ll_0$&$g$&$\kk_{\mbox{e}}$&$(\ll_{\mbox{e}}/\ll_0)$\\ \hline\hline
128&1.0&.05&.722(2)&.726(2)\\
   &   &   &(.717)&(.717)\\\hline
256&1.5&.05&.470(5)&.474(2)\\ 
   &   &   &(.472)&(.472)\\\hline
256&2.0&.05&.269(4)&.266(11)\\
   &   &   &(.264)&(.264)\\\hline
512&2.5&.05&.139(5)&.140(20)\\
   &   &   &(.125)&(.125)\\\hline
\end{tabular} }
\caption{$N$ {\it is the number of modes in random field. Theoretical values in brackets.}}
\end{table}
\vskip 5 truemm

which is based on a continuum construction for $\phi(\bx)$, has been described 
in previous papers \cite{USDIF1,USDIF2,USDIF3}. An important point for the present
paper is that the random field is constructed from a set of $N$ modes and the integrity
of the Gaussian property of the statistics of the random field is dependent 
on having a sufficiently large value of $N$~.

The results clearly 
show the equality of the two renormalization 
parameters for a wide range of disorder parameter. This common value is also
equal, as was found in previous work, to the RGC prediction of 
$\exp{\{-{1\over 3}{\ll_0^2/\kk_0^2}\}}$. There is a slight discrepancy
at the higher values of the disorder parameter. We feel that
this is a systematic error in the simulation due to the limited number of
modes incorporated in the random field. Another possible source of error is 
that the value of the drift parameter has become
so large that $O(g^2)$ effects are influencing the values
of the measured quantities. It is also possible that 
the assumptions behind the renormalization group 
calculation may no longer be valid at these values of $\ll_0$~. 
It is interesting to note that, nevertheless the equality
of the two renormalization factors is maintained throughout with particular
accuracy. We can be reasonably confident therefore
of our renormalization group results and the associated Ward 
identity \cite{USDIF2,USDIF3} in this isotropic case.

\section{``Ward'' Identity}

In previous work we suggested a Ward identity as an explanation of the 
proportionality of the renormalization of the vertex and diffusivity
matrix. The identity was verified to two-loop order in perturbation theory.
Here we show that this Ward identity changes form when the bare vertex 
and diffusivity matrices are no longer proportional to one another. We will only discuss
the the change at one-loop order since this is sufficient to demonstrate the breakdown.

From the formulae of the previous section we see that to one-loop 
\beq
\fr{\d}{\d k_i}\SS(\bk)=[\kk^0(\ll^0)^{-1}]_{ij}V_j(\bk,\bk)+U_i(\bk)~~,
\eeq
where
\beq
U_i(\bk)=-\int\dmq D(\bq)\fr{q_r\ll^0_{ri}q_m\ll^0_{mn}k_n+q_r\ll^0_{rs}(\bq+\bk)_sq_m\ll^0_{mi}}
                                               {(\bq+\bk)_j\kk^0_{jl}(\bq+\bk)_l}~~.
\eeq
For small $\bk$ we see that
\beq
k_iU_i(\bk)=-\int\dmq D(\bq)\left\{\fr{2(q_s\ll^0_{rs}k_s)^2}{q_m\kk^0_{mn}q_n}
                                 -\fr{2q_r\ll^0_{rs}q_s~q_n\ll^0_{mn}k_n~q_j\kk^0_{jl}k_l}
                                      {(q_m\kk^0_{mn}q_n)^2}\right\}
                                                +O(k^4)~~.
\label{XX}
\eeq
It is easy to see that that the right side of this equation vanishes to $O(k^4)$
when $\kk^0\propto \ll^0$~. In general however it does not do so. A simple case
to consider is one in which the statistics of the $\phi(\bx)$-field are isotropic, 
the diffusivity has the form $\kk^0_{ij}=\kk_0\delta_{ij}$ but the drift coefficient
retains an anisotropic tensorial structure. We easily evaluate the right 
side of eq(\ref{XX}) to be
\beq
k_iU_i(\bk)=-\fr{2\DD(0)}{3\kk_0}\left\{\fr{3}{5}k_i[(\ll^0)^2]_{ij}k_j
                                 -\fr{1}{5}\ll^0_{mm}k_i\ll^0_{ij}k_j\right\}
                                             +O(k^4)~~.
\eeq
If we define
\beq
\bar{\ll}^0_{ij}=\ll^0_{ij}-\fr{1}{3}\ll^0_{mm}~~,
\eeq
so that $\bar{\ll}^0$ is the traceless or quadrupole part of $\ll^0$ then we can recast
the above equation in the form
\beq
k_iU_i(\bk)=-\fr{2\DD(0)}{3\kk_0}\left\{\fr{3}{5}k_i[(\bar{\ll}^0)^2]_{ij}k_j
                                 +\fr{1}{5}\ll^0_{mm}k_i\bar{\ll}^0_{ij}k_j\right\}
                                                 +O(k^4)~~,
\eeq
that is
\beq
k_iU_i(\bk)=-\fr{2\DD(0)}{5\kk_0}k_i[\bar{\ll}^0\ll^0]_{ij}k_j+O(k^4)~~.
\eeq
This form of the result makes it clear that when the traceless part of the drift
tensor $\bar{\ll}^0$ vanishes we return to the situation in which the original
Ward identity holds. 

For the particular case we are considering the the modified Ward identity
implies for small $\bk$ the result
\beq
\fr{\d}{\d k_i}\SS(\bk)=-\fr{\DD(0)}{15\kk_0}\left\{\ll^0_{mm}\ll^0_{ij}+2\ll^0_{il}\ll^0_{lj}\right\}k_j
  -\fr{2\DD(0)}{3\kk_0}\left\{\fr{3}{5}[(\ll^0)^2]_{ij}k_j
                                 -\fr{1}{5}\ll^0_{mm}\ll^0_{ij}k_j\right\}
                                             +O(k^3)~~.
\eeq
On the basis of this calculation we do not expect a simple relationship between 
the macroscopic diffusivity and the macroscopic drift coefficient. This is 
confirmed in the next section by a renormalization group calculation.

\section{Renormalization Group Equations}

The breakdown of the Ward identity and the absence of any simple relation between
the effective diffusivity and drift tensors makes it
interesting to examine the consequences of the renormalization group
approach to computing the macroscopic parameters. These equations for the
diffusivity tensor were written down in refs \cite{USDIF2,USDIF3}. We complete the scheme here 
with the corresponding equation for the drift coefficient. For simplicity we 
write down the equations for the wavenumber slicing scheme. They are
\beq
\fr{d\kk_{ij}}{d\Lambda}=\int\fr{d^3\bq}{(2\pi)^3}\delta(q-\Lambda)D(q)
                            \fr{(q_m\kk_{mi}q_n\ll_{nj}+q_m\kk_{mj}q_n\ll_{ni})q_r\ll_{rs}q_s
                                      -q_m\ll_{mi}q_n\ll_{nj}q_r\kk_{rs}q_s}{(q_r\kk_{rs}q_s)^2}
\label{RGE1}
\eeq
and
\beq
\fr{d\ll_{ij}}{d\Lambda}=\int\fr{d^3\bq}{(2\pi)^3}\delta(q-\Lambda)D(q)
                                      \fr{q_m\ll_{mi}q_n\ll_{nj}q_r\ll_{rs}q_s}{(q_r\kk_{rs}q_s)^2}
\label{RGE2}
\eeq
It is clear from these equations that, as mentioned in a previous paper \cite{USDIF3},
those solutions satisfying boundary conditions for which
$\ll^0_{ij}\propto \kk^0_{ij}$ maintain the the proportionality 
$\ll_{ij}(\Lambda)\propto \kk_{ij}(\Lambda)$
for all $\Lambda$ with a constant ratio.
 
It is easily checked that of course the isotropic solutions are
those of earlier papers, namely
\beq
\kk_{ij}(\Lambda)=\kk^S_{ij}(\Lambda)
        =\kk_0\exp\left\{-\fr{1}{3}\fr{\ll^2_0}{\kk^2_0}\Delta_{\Lambda}(0)\right\}\delta_{ij}
\eeq
and
\beq
\ll_{ij}(\Lambda)=\ll^S_{ij}(\Lambda)
          =\ll_0\exp\left\{-\fr{1}{3}\fr{\ll^2_0}{\kk^2_0}\Delta_{\Lambda}(0)\right\}\delta_{ij}
\eeq
where
\beq
\Delta_{\Lambda}(0)=\int_{q>\Lambda}\fr{d^3\bq}{(2\pi)^3}D(q)~~.
\eeq

We can examine solutions nearby these isotropic solutions, that are perturbed by a 
small anisotropic change in the local drift tensor. It is no restriction to make this
small change traceless. We have then
\begin{eqnarray}
\kk_{ij}(\Lambda)&=&\kk^S_{ij}(\Lambda)+\eta_{ij}\\ \nonumber
\ll_{ij}(\Lambda)&=&\ll^S_{ij}(\Lambda)+\mu_{ij}
\end{eqnarray}
where
\begin{eqnarray}
\eta_{ij}(\infty)&=&0\\ \nonumber
\mu_{ij}(\infty)&=&\mu^0_{ij}
\end{eqnarray}
with $\mu^0_{ii}=0$~.
A perturbative analysis of eqs(\ref{RGE1},\ref{RGE2}) yields
\begin{eqnarray}
\fr{d\eta_{ij}}{d\Lambda}&=&\fr{\ll_0}{\kk_0}\int\dmq\delta(q-\Lambda)D(q)\left\{\fr{\ll_0}{\kk_0}\fr{2}{3}\eta_{ij}
                                       -\fr{3}{15}\fr{\ll_0}{\kk_0}(\eta\delta_{ij}+2\eta_{ij})\right.\\ \nonumber
       &&~~~~~~~~~~~~~~~~~~~~~~~~~~~~~~~~~~~~~~~~~~~~~~~~~~~~~~~~
                                    \left.+\fr{2}{15}(\mu\delta_{ij}+2\mu_{ij})\right\}\\ \nonumber
\fr{d\mu_{ij}}{d\Lambda}&=&\fr{\ll_0^2}{\kk_0^2}\int\dmq\delta(q-\Lambda)D(q)\left\{\fr{2}{3}\mu_{ij}
                                       +\fr{1}{15}(\mu\delta_{ij}+2\mu_{ij})
                                       -\fr{2}{15}\fr{\ll_0}{\kk_0}(\eta\delta_{ij}+2\eta_{ij})\right\}
\end{eqnarray} 
If we introduce a variable $0<s<1$ such that
\beq
s=\int_{q<\Lambda}\dmq D(q)
\eeq
and impose the allowed constraint that $\eta=\eta_{ii}=0$ and $\mu=\mu_{ii}=0$ we find
\begin{eqnarray}
\fr{d\eta_{ij}}{d~s}&=&\fr{4}{15}\fr{\ll_0^2}{\kk_0^2}\eta_{ij}+\fr{4}{15}\fr{\ll_0}{\kk_0}\mu_{ij}\\ \nonumber
\fr{d\mu_{ij}}{d~s}&=&\fr{12}{15}\fr{\ll_0^2}{\kk_0^2}\mu_{ij}-\fr{4}{15}\fr{\ll_0^3}{\kk_0^3}\eta_{ij}
\end{eqnarray}
These equations are easily integrated  and yield the result
\begin{eqnarray}
\kk_{ij}&=&\kk_0\exp\left\{-\fr{1}{3}\fr{\ll_0^2}{\kk_0^2}\right\}\delta_{ij}
                   -\fr{4}{15}\fr{\ll_0}{\kk_0}\exp\left\{-\fr{8}{15}\fr{\ll_0^2}{\kk_0^2}\right\}
                                                                    \mu^0_{ij}\\ \nonumber
\ll_{ij}&=&\ll_0\exp\left\{-\fr{1}{3}\fr{\ll_0^2}{\kk_0^2}\right\}\delta_{ij}
                               +\left(1-\fr{4}{15}\fr{\ll_0^2}{\kk_0^2}\right)
                          \exp\left\{-\fr{8}{15}\fr{\ll_0^2}{\kk_0^2}\right\}\mu^0_{ij}
\label{NONSR}
\end{eqnarray}
It follows that 
\beq
(\ll\kk^{-1})_{ij}=\fr{\ll_0}{\kk_0}\left(\delta_{ij}
    +\fr{1}{\ll_0}\exp\left\{-\fr{1}{5}\fr{\ll_0^2}{\kk_0^2}\right\}\mu^0_{ij}\right)~~.
\eeq
For this near isotropic case at least we see that the proportionality of the
macroscopic drift and diffusivity tensors is restored exponentially as the
renormalization process takes place.

\section{Numerical Simulation in the Anisotropic Case}

We tested the above results from the renormalization group calculation
against a simulation for which the asymmetric local drift tensor
is diagonal in the coordinate basis with elements 
\beq
\mu^{0}=\left(\begin{array}{ccc}
                 -\alpha/2&0&0\\
                  0&-\alpha/2&0\\
                  0&0&\alpha
                   \end{array}\right)~~.
\eeq
Because of the axial symmetry of the local drift tensor the same
property ensures that the effective diffusion and drift tensors will
have the form 
\beq
\kk_{\mbox{e}}=\left(\begin{array}{ccc}
                   \kk_{\perp}&0&0\\
                    0&\kk_{\perp}&0\\
                    0&0&\kk_{\parallel}
                      \end{array}\right)
                          ~~~~\mbox{and}~~~~\ll_{\mbox{e}}=\left(\begin{array}{ccc}
                                                                 \ll_{\perp}&0&0\\
                                                                 0&\ll_{\perp}&0\\
                                                                 0&0&\ll_{\parallel}
                                                                   \end{array}\right)
\eeq
The results of the numerical simulation for certain values of $\ll_0$ and 
various values of $\alpha$ are shown  in Table 2. 

\begin{table}
\center
\begin{tabular}{|c|c|c|c|c|c|}\hline
$\ll_0$&$\alpha$&$\kk_{\parallel}$&$\kk_{\perp}$
                &$\ll_{\parallel}$&$\ll_{\perp}$\\\hline\hline
1.&.2&.701(3)&.735(3)&.798(5)&.660(5)\\
  &  &(.685)& (.732)& (.803)&(.674)\\\hline
1.&.1&.698(3)&.727(2)&.770(6)&.687(5)\\
  &  &(.701)&(.724)&(.760)&(.695)\\\hline
1.&-.1&.730(5)&.712(2)&.672(5)&.730(5)\\
  &   &(.732)&(.709)&(.673)&(.737)\\\hline
1.&-.2&.764(4)&.701(3)&.622(6)&.772(7)\\
  &   &(.748)&(.701)&(.630)&(.760)\\\hline
\end{tabular}
\caption{{\it Number of modes in random field is} 256. {\it Theoretical values
are shown in brackets.}}
\end{table}
\vskip 5 truemm 
The results with some small discrepancies for the two cases with
$|\alpha|=2$ are in good agreement with the predictions of the the
renormalization group calculations expressed in eqs(\ref{NONSR})~.
It is not unreasonable that an asymmetry parameter as large as 20\%
is the limit of applicability of the simple perturbation approach in the
previous section. To check the results by simulation in finer detail for 
smaller values of $|\alpha|$ requires higher statistical accuracy 
than can easily be achieved. Our simulations typically involved 256
particles in 256 velocity fields and required 100/200 processor hours.
Nevertheless we can conclude with reasonable confidence that the RGC
has produces an accurate result even in the asymmetric case.
It may be that an important condition for the success of the RGC is
the isotropy of the random field statistics.

\section{Conclusions}

In this paper we have confirmed the importance of the role of drift in the set
of effective parameters that govern the long time behaviour of a diffusion 
process in which a particle moves subject to molecular diffusion and 
to the influence of the gradient of a random scalar field. 
Another way of expressing this is to say that in addition to the the
diffusion process itself, it is important to consider the effect on the system 
of long range external fields and the strength with which they are
coupled to the system. We have exploited this effect by computing and
measuring the mean drift in a simulation induced by an external field
of constant gradient. The effects however must also be relevant to all
external fields of long range. The conclusion, arrived at in previous work \cite{mythesis,dc,
USDIF1,USDIF2}, and confirmed in this simulation is that the effective 
drift and diffusivity parameters are renormalized relative to their
local values by identical factors in the isotropic case.

We showed how the renormalization of the drift parameter could be
interpreted as a result of biased {\it spatial} averaging effects
due to the density distribution adopted by particles passing through
the medium represented by the random field under the influence of an 
external field of constant gradient.

We also examined a situation in which the isotropy is broken by giving the drift
tensor an axi-symmetric form. We confirm that the Ward identity suggested
in previous work as an explanation for the equality of the diffusion and drift 
renormalization factors is indeed broken in lowest order perturbation theory.
We do not expect a simple relationship between effective diffusion and drift
in this case. This is confirmed at a theoretical level by using the
renormalization group approach in a near symmetric situation. The predictions
of the theoretical calculation are quite well verified by the results
of a simulation. The implication is that the renormalization group calculation
will work reasonably well when the statistics of the random scalar field are 
isotropic.

\section*{Acknowldgements}

This work was completed with the support from EU Grant CHRX-CT93-0411.
The calculations were performed on the Hitachi SR2001 located at 
the headquarters of Hitachi Europe Ltd, Maidenhead, Berkshire, England.

\pagestyle{empty}
\newpage
\begin{figure}[htb]
\begin{center}\leavevmode
\epsfxsize=7 truecm\epsfbox{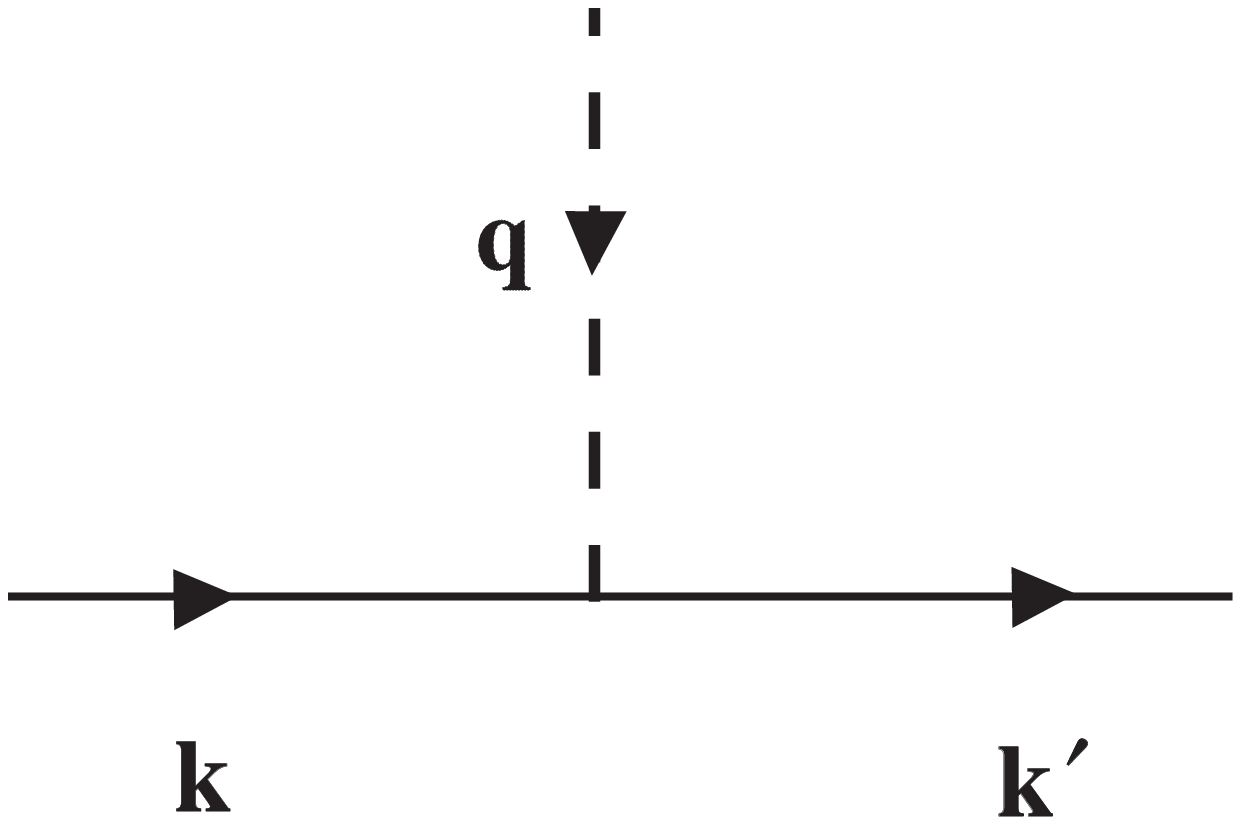}
\end{center}
\caption[]{Vertex diagram}
\label{figure:Fig1}
\end{figure}

\begin{figure}[htb]
\begin{center}\leavevmode
\epsfxsize=7 truecm\epsfbox{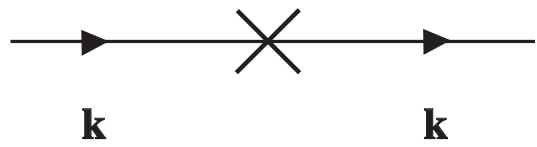}
\end{center}
\caption[]{ Drift insertion vertex}
\label{figure:Fig2}
\end{figure}

\begin{figure}[htb]
\begin{center}\leavevmode
\epsfxsize=7 truecm\epsfbox{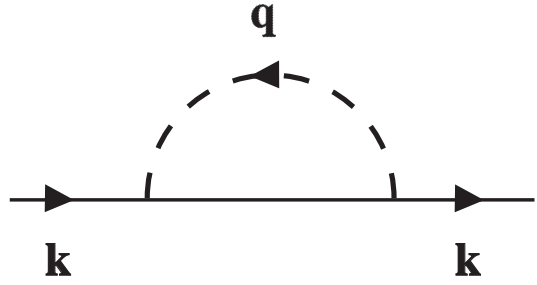}
\end{center}
\caption[]{ One--loop contribution to $\Sigma$}
\label{figure:Fig3}
\end{figure}

\begin{figure}[htb]
\begin{center}\leavevmode
\epsfxsize=7 truecm\epsfbox{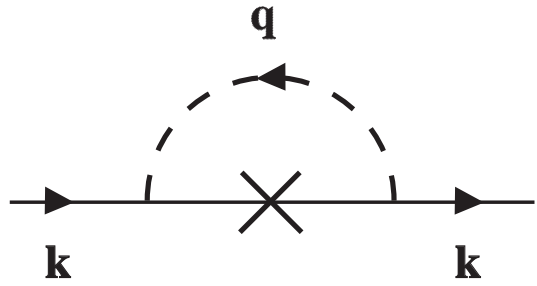}
\end{center}
\caption[]{ One-loop correction to drift insertion}
\label{figure:Fig4}
\end{figure}

\begin{figure}[htb]
\begin{center}\leavevmode
\epsfxsize=7 truecm\epsfbox{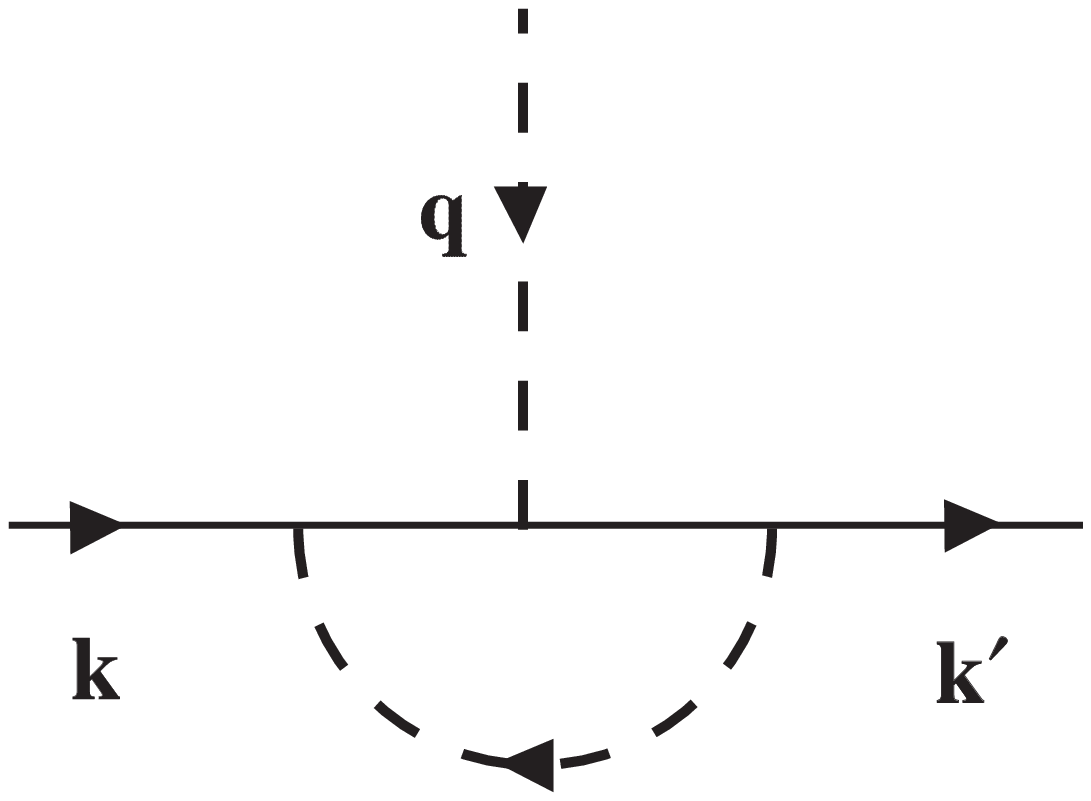}
\end{center}
\caption[]{One loop correction to vertex}
\label{figure:Fig5}
\end{figure}

\newpage
%

\newpage

\begin{thebibliography}{1}

%
%
%
\bibitem [1]{mythesis}
D.S. Dean 
{\em Stochastic Dynamics}, PhD Thesis, University of Cambridge, 1993.

\bibitem [2]{dc}
M W Deem and D Chandler
{\em Journal of Stat. Phys.}, {\bf 76}, 911, 1994

\bibitem [3]{USDIF1}
D.S. Dean, I.T. Drummond and R.R. Horgan,
{\em J. Phys. A: Math Gen.}, {\bf 27}, 5135, 1994.

\bibitem [4]{USDIF2}
D. S. Dean, I. T. Drummond and R. R. Horgan,
{\em Journal of Physics A: Math. Gen.}, {\bf 28} 1235-1242, 1995.

\bibitem[5]{USDIF3}
D. S. Dean, I. T. Drummond and R. R. Horgan
{\em J. Phys. A: Math.Gen.}, {\bf 28} 6013-6025, (1995)

%


\end{thebibliography}
\end{document}